\documentclass[11pt,notitlepage,aas,nofootinbib,superscriptaddress,floatfix,twocolumn]{aastex631}

\usepackage{float}
\usepackage{amsmath,bm}
\usepackage{amsfonts}
\usepackage{graphicx}
\usepackage{multirow}
\usepackage{aas_macros}
\usepackage[title]{appendix}
\usepackage{amssymb}
\usepackage{array, longtable}
\usepackage{orcidlink}

\definecolor{mpl_blue}{HTML}{1F77B4}
\definecolor{mpl_orange}{HTML}{FF7F0E}
\definecolor{mpl_green}{HTML}{2CA02C}
\definecolor{mpl_red}{HTML}{D62728}

\newcommand{\enterprise}{\texttt{enterprise}}
\newcommand{\entext}{\texttt{enterprise\_extensions}}
\newcommand{\ptmcmc}{\texttt{PTMCMCSampler}}
\newcommand{\holodeck}{\texttt{holodeck}}

\newcommand{\Agw}{\ensuremath{A_\mathrm{gw}}}
\newcommand{\gammagw}{\ensuremath{\gamma_\mathrm{gw}}}

\begin{document}

\title{Characterizing the nanohertz gravitational wave background using a $t$-process power spectral density}

\author[0009-0006-5476-3603]{Shashwat C.~Sardesai}
\affiliation{Center for Gravitation, Cosmology and Astrophysics \\
University of Wisconsin--Milwaukee \\
P.O. Box 413, Milwaukee WI, 53201, USA}

\author[0000-0003-1407-6607]{Joseph Simon}
\altaffiliation{NSF Astronomy and Astrophysics Postdoctoral Fellow}
\affiliation{Department of Astrophysical and Planetary Sciences \\
University of Colorado, Boulder \\
CO 80309, USA}

\author[0000-0003-4700-9072]{Sarah J.~Vigeland}
\affiliation{Center for Gravitation, Cosmology and Astrophysics \\
University of Wisconsin--Milwaukee \\
P.O. Box 413, Milwaukee WI, 53201, USA}

\begin{abstract}
Pulsar timing arrays are sensitive to low-frequency gravitational waves (GWs), which induce correlated changes in millisecond pulsars' timing residuals. PTA collaborations around the world have recently announced 
evidence of a nanohertz gravitational wave background (GWB), which may be produced by a population of supermassive black hole binaries (SMBHBs). The GWB is often modeled as following a power-law power spectral density (PSD); however, a GWB produced by a cosmological population of SMBHBs is expected to have a more complex power spectrum due to the discrete nature of the sources. In this paper, we investigate using a $t$-process PSD to model the GWB, which allows us to fit for both the underlying power-law amplitude and spectral index as well as deviations from that power law, which may be produced by individual nearby binaries. 
We create simulated data sets based on the properties of the 
NANOGrav 15-year data set, and we demonstrate that the $t$-process PSD can accurately recover the PSD when deviations from a power-law are present. With longer timed datasets and more pulsars, we expect the sensitivity of our PTAs to improve, which will allow us to precisely measure the PSD of the GWB and study the sources producing it.
\end{abstract}


\section{Introduction}

Massive objects orbiting each other radiate away energy in the form of gravitational waves (GWs).
Supermassive black hole binaries (SMBHBs), 
which form following massive galaxy mergers, 
emit GWs with frequencies ranging from $10^{-9} - 10^{-7}$ Hz when they are 
at $\sim10^{-1}-10^{-2}$ pc separations \citep{2013CQGra..30x4009S, 2019A&ARv..27....5B}. 
Pulsar timing arrays (PTAs) detect GWs at nanohertz frequencies 
by looking for the time delay induced in the pulses of a collection of rapidly spinning millisecond pulsars (MSPs) \citep{1978SvA....22...36S, 1979ApJ...234.1100D, 1983ApJ...265L..39H}. 
MSPs are recycled neutron stars with extremely low spin down rates, and their rotational stability makes them ideal for GW detection \citep{1990ApJ...361..300F}. 

The superposition of GWs emitted by a cosmological population of 
SMBHBs is expected to give rise to a stochastic gravitational wave background (GWB) \citep{1995ApJ...446..543R, 2003ApJ...583..616J, 2003ApJ...590..691W, 2015MNRAS.451.2417R}. 
Recently, the North American Nanohertz Observatory for Gravitational Waves (NANOGrav), the European Pulsar Timing Array (EPTA), the Indian Pulsar Timing Array (InPTA), the Parkes Pulsar Timing Array (PPTA), and the Chinese Pulsar Timing Array (CPTA) announced the first evidence for a GWB at nanohertz frequencies 
\citep{2023ApJ...951L...8A, 2023A&A...678A..50E, 2023ApJ...951L...6R, 2023RAA....23g5024X}. 
The observed signal is consistent with a GWB produced 
by SMBHBs \citep{2023ApJ...952L..37A,2024A&A...685A..94E}, but is also consistent with a GWB produced by other sources such as inflation, phase transitions, and cosmic strings (e.g., \citet{2023ApJ...951L..11A}). 
An analysis by the International Pulsar Timing Array (IPTA) found that the results from NANOGrav, the EPTA, the InPTA, and the PPTA are broadly consistent with one another \citep{2024ApJ...966..105A}, 
and work is underway preparing the third IPTA data set that will combine these data with data from the MeerKat Pulsar Timing Array \citep{2023MNRAS.519.3976M}, the Canadian Hydrogen Intensity Mapping Experiment (CHIME) \citep{2021ApJS..255....5C}, and the Low-Frequency Array (LOFAR) \citep{2013A&A...556A...2V}. 
With longer timing baselines and more pulsars, the sensitivity of PTAs will increase, allowing for more precise measurements of the GWB and the study of its astrophysical or cosmological sources \citep{2013CQGra..30v4015S}.

For a population of circular SMBHBs evolving only due to GW emission, the GWB power spectral density (PSD) is expected to follow, $P(f) \propto f^{-13/3}$ \citep{2001astro.ph..8028P}.
However, the PSD for an astrophysically realistic GWB is expected to be more complicated.
Environmental effects such as stellar hardening and alpha-disk interactions accelerate the orbital evolution at large separations, resulting in different power-law spectral indices at the low-frequency and high-frequency ends of the spectrum \citep{2011MNRAS.411.1467K, 2003AIPC..686..201M, 2015PhRvD..91h4055S}, 
while the presence of a significant fraction of SMBHBs with 
non-negligible eccentricity will result in a flattening of the 
GWB spectrum \citep{2020PhRvD.101d3022S, 2024MNRAS.532..295F}. 
Because of the discrete nature of the sources, 
we also expect to find deviations from a power-law PSD 
at higher frequencies where fewer sources are contributing 
to the GWB. There may also be excess power at individual frequencies caused by a small number of loud, nearby sources \citep{2022ApJ...941..119B,2024arXiv240407020A}. The finiteness of the population of SMBHBs can lead to spectral variation in the GWB as shown in \citet{2024ApJ...971L..10L}.

In this paper, we explore the validity of detecting deviations from a power law PSD. We create simulations of the NANOGrav 15yr datasets, consisting of 67 pulsars, and employ a Bayesian analysis to fit multiple PSD models. We introduce a ``$t$-process'' model, which consists of an underlying power law convolved with multiplicative variables, each described by an inverse gamma distribution prior, at each frequency and compare the findings to other standard PSD models used in PTA analyses.

This paper is organized as follows. In Section \ref{methods} we give an overview of PTAs, we discuss the $t$-process model as well as the other models used to describe the GWB PSD, and we describe our analysis methods. In Section \ref{sims} we present the results of our analysis, comparing the results of the $t$-process to other PSD models across different simulations containing: a pure power-law GWB, a GWB with excessive noise at a single frequency component, and a simulated astrophysical background from SMBHBs.
Finally, in Section \ref{concl} we summarize our results and look to the future.

\section{Model and Methods} \label{methods}

\subsection{PTA overview}

Pulsar timing residuals are the difference between the observed times of arrival and the expected, deterministic times of arrival. It can also be represented as a combination of noise processes

\begin{equation}
    \delta t = M \epsilon + F a + n \,,
\end{equation}

\noindent where $M \epsilon$ are linear perturbations to the timing model, $Fa$ is the the contribution from intrinsic red noise as well as the common process (GWB), and $n$ is the contribution of white noise. The intrinsic red noise can be modeled as a power law with an amplitude and a spectral index.

\begin{eqnarray}
    P_{\mathrm{rn}}(f) &=& A_{\mathrm{rn}}^2 \left(\frac{f}{f_{\mathrm{yr}}}\right)^{-\gamma} \mathrm{yr^3} . 
\end{eqnarray}

Each pulsar is expected to be affected by intrinsic red noise processes (e.g., spin noise \citep{2010ApJ...725.1607S, 2017ApJ...834...35L, 10.1093/mnras/staa615}, variations in the interstellar medium \citep{2017ApJ...834...35L, 2017ApJ...841..125J}), white noise processes (e.g., radiometer noise, pulse jitter \citep{ 1985ApJS...59..343C, 2016ApJ...819..155L, 2019ApJ...872..193L, 2021MNRAS.502..407P}), and the GW background itself.
It has been shown that in some cases that intrinsic red noise can be mistaken for a common process \citep{2022ApJ...932L..22G, 2021ApJ...917L..19G, 2022MNRAS.516..410Z, 2024ApJS..273...23V} and is very dependent on the Bayesian priors chosen. However, we can distinguish between intrinisic red noise and the GWB based on the correlations between pairs of pulsars -- the GWB induces interpulsar correlations that follow the Hellings-Downs curve \citep{1983ApJ...265L..39H}, while noise processes will be uncorrelated or induce other types of correlations \citep{2016MNRAS.455.4339T}. 

\subsection{Common red noise signal}

The incoherent superposition of GWs produced by a population 
of SMBHBs leads to a GW background that can be described in terms of its characteristic strain $h_c(f)$. 
For a population of circular SMBHBs evolving only due to GW emission, this is predicted to follow a power-law \citep{2001astro.ph..8028P}, 

\begin{equation} \label{char strain}
    h_c = \Agw \left( \frac{f}{f_{\mathrm{yr}}} \right)^{-2/3} ,
\end{equation}

\noindent where $f$ is the GW frequency and $A$ is the amplitude at a reference frequency of $f_{\mathrm{yr}} = 1 \; \mathrm{yr}^{-1} = 31.7\; \mathrm{nHz}$. It is common to use a reference frequency of $1 \mathrm{yr}^{-1}$, but other reference frequencies such as $0.1 \mathrm{yr}^{-1}$ may be used.

PTAs detect GWs by measuring the redshift induced in the pulses of a collection of millisecond pulsars. 
The response in a PTA is a time delay induced in the pulse arrival times of all the pulsars.
The measured residual power spectral density is related to the characteristic strain spectrum according to

\begin{eqnarray} \label{residual psd}
    P_{\mathrm{pl}}(f) &=& \frac{h_c^2(f)}{12 \pi^2 f^3} df \\ \nonumber
    &=& \frac{\Agw^2}{12 \pi^2} \left( \frac{f}{f_\mathrm{yr}} \right)^{-\gammagw} df,
\end{eqnarray}

\noindent where $\Agw$ is the amplitude of the characteristic strain from Eq.~\eqref{char strain}, $df = 1/T_\mathrm{span}$, the total time span of the PTA, and $\gammagw$ is the spectral index in the power spectral density. For purely circular SMBHBs, $\gammagw = 13/3$.

On average, we expect the GW background to produce a residual power spectrum according to Eq.~\eqref{residual psd}. However, an astrophysically realistic spectrum will deviate from a power-law PSD due to the finite number of binaries contributing to the spectrum \citep{2008MNRAS.390..192S}. There also may be a turnover in the spectrum at low frequencies if interactions with the environment extend into the PTA band \citep{2015PhRvD..91h4055S}. At high frequencies, the power spectrum of the residuals flattens due to white noise. 

We model the red noise processes using a Fourier series with linearly spaced frequencies $f_i = i/T_\mathrm{span}$, where $T_\mathrm{span}$ is the total span of the data and $i$ is an integer. For our analyses we use $i = 1, 2, \ldots, 30$, which corresponds to a frequency range of $1.98$ to $59.3 \; \mathrm{nHz}$.
To capture the subtleties at each frequency component, one can make use of a free spectral approach, which treats the power at each frequency as an independent parameter,

\begin{equation}
    P_{\mathrm{fs}}(f_{i}) = \frac{h_c^2(f_i)}{12 \pi^2 f_i^3 } df = \delta t^{2}_{\mathrm{delay}} (f_i) \,. \label{Free spectral}
\end{equation}
By design, there is no relation between the power at different frequencies: the free spectral is completely model agnostic and purely measures the power at each frequency individually.

The $t$-process model is a modification to the power law where every frequency component has a multiplicative factor governed by an inverse gamma distribution prior, 
\begin{eqnarray} \label{t-process}
    P_{tp} (f_{i}) &=& \alpha_{i} P_\mathrm{pl}(f_i) \nonumber \\ 
    \alpha_{i} &\sim& \mathrm{invgam}(1,1) , \
\end{eqnarray}

\noindent The inverse gamma function follows a probability of
\begin{equation}
    \mathrm{invgam}(\alpha_i, 1,1) = \frac{1}{\Gamma(1)} \alpha_i^{-2} \exp{\left(-\frac{1}{\alpha_i}\right)}
\end{equation}

\noindent where $P_\mathrm{pl}(f_i)$ is the standard power law model from Eq. ~\eqref{residual psd}, and $\alpha_{i}$ is a multiplicative factor that adjusts to power at a frequency $f_i$ compared to the underlying power-law PSD.
This model is a good compromise between the power-law model, which assumes a relationship between the power at different frequencies, and the agnostic free spectral model 
since it allows for deviations in the PSD from a power-law. 
This model was previously used to describe the intrinsic red noise of pulsar J0613$-$0200 in \citet{2020ApJ...900..102A} 
since that pulsar showed both low-frequency red noise 
and excess noise at $f_\mathrm{gw} = 15\;\mathrm{nHz}$. 
In this work, we use it to model the GWB because it is able to model the underlying power-law PSD of the GWB, as well as capture and quantify deviations from the power-law.

\subsection{Bayesian Methods and Software}

We create simulations of the NANOGrav 15yr dataset using the \enterprise\ \citep{2019ascl.soft12015E} and \entext\ \citep{enterprise} packages. Our likelihood is modeled as a Gaussian,

\begin{equation}
    p(\Vec{\delta t} | \phi) = \frac{\mathrm{exp}\left( - \frac{1}{2} \Vec{\delta t}^{T} \Sigma^{-1} \Vec{\delta t} \right)}{\sqrt{2 \pi (\mathrm{det}\Sigma)}} ,
\end{equation}

\noindent where $\Vec{\delta t}$ is the vector of residuals of all pulsars, $\phi$ is the vector of parameters drawn from the priors, and $\Sigma$ is the covariance block matrix of all pulsars.
We use \ptmcmc\ \citep{2019ascl.soft12017E} to draw values from the posteriors for 
intrinsic red noise and common red noise parameters using a Markov Chain Monte Carlo (MCMC) approach. For the purposes of this paper, interstellar variations/dispersion medium, pulse jitter, and correlated white noise are not considered.

\subsection{Simulations}

Our simulated PTA is based on the NANOGrav 15yr dataset \citet{2023ApJ...951L...8A} which contains 67 pulsars and a baseline of 16.03 yrs. In each of our simulations, the injected uncertainty in TOAs, and intrinsic pulsar red noise are equal to the measured quantities in the pulsars in the 15yr data set. This is done by running a Bayesian analysis with intrinsic red noise included for each pulsar as well as a common red noise process across the entire array. We then only consider the maximum likelihood intrinsic red noise values (this is to prevent contamination of the pulsar-dependent red noise with the common red noise signal present in the dataset). We then inject a new common red noise spectra using \enterprise\ and analyze the results.

To test the $t$-process, we use three different types of injected common signal. We use (1) a pure power-law, (2) a power-law with a known amount of excess noise at $5.93$~nHz (the third frequency bin),
and finally (3) an astrophysically realistic background from a simulated population of SMBHBs created by the \holodeck\footnote{ https://github.com/nanograv/holodeck}\citep{2023ApJ...952L..37A} simulation package. To create a realistic simulation of the GWB, we must take into account the evolution and merging of galaxies that host the SMBHs. The \holodeck\ package uses a semi-analytical model, which includes hyper-parameters for galactic merger time, galaxy stellar mass function, galaxy close pair fraction, and SMBH mass - host galaxy relation to get a differential number density of SMBHBs \citep{2017PhDT.......206S}. 
The characteristic strain from a population of SMBHBs is then calculated by \cite{2012arXiv1211.4590M, 2001astro.ph..8028P, 2008MNRAS.390..192S, 2015PhRvD..91h4055S}
\begin{equation}
    h_{c}^2 (f) = \int dz \int dM \int dq \frac{d^4 N}{dM \, dq \, dz \, d\ln{f}} h_{s}^2 (f)   . \\
\end{equation}
Here, $N$ is the total number of SMBHBs present within each bin of total mass $M$, mass ratio $q$, redshift $z$, and log frequency $\ln f$.
Since we are dealing with discrete sources, we must modify our integral to be a sum. The \holodeck\ package uses a Poisson distribution to estimate the number of binaries within a range of $dz$, $dM$ and $dq$ at each $d\ln{f}$ to calculate the overall characteristic strain from individual sources as shown in \citet{2017PhDT.......206S, 2023ApJ...952L..37A}
\begin{equation}
    h_{c}^2 (f) = \sum_{M, q, z, f}  \mathcal{P} 
    \left( \frac{\partial^4 N}{\partial M \partial q \partial z \partial \ln{f}} \Delta M \Delta q \Delta z \Delta \ln{f} \right) 
    \frac{h_s^2(f)}{\Delta \ln{f}} , \\
\end{equation}
where $\mathcal{P}$ indicates a random number drawn from a Poisson distribution with the mean value given by the number of binaries in each bin of $M$, $q$, $z$, and $\ln{f}$. For a large population of sources, $h_c^2$ closely follows a power-law, as described in Eq.~\eqref{char strain}. However, when there the signal is dominated by a small number of discrete sources, there is a breakdown of stochasticity and a subsequent deviation from power-law behavior: this can be due to the signal being dominated by a single loud binary, or because the binaries are evolving quickly at high GW frequency.

\section{Results} \label{sims}

For all of our simulations, we use three models to reconstruct the PSD: a power-law, a $t$-process, and a free spectrum. 
We show reconstructions of the PSDs using all three models and compare to the injected GWB. 
For the power-law and $t$-process models, we also show corner plots for the model parameter posteriors. 
This allows us to compare the recovered parameter posteriors to the injected values, as well as explore the covariances between model parameters.

For the $t$-process model, the $\alpha$ parameters provide a way to measure deviations from a power-law PSD. If there are no deviations present, the posteriors for $\alpha$ will peak at 1. In order to assess the significance of deviations, we use the Hellinger distance \citep{hellinger} to compare the posteriors and priors. The Hellinger distance is a measure of the similarity between two distributions: a Hellinger distance of 0 means the two distributions are the same and 1 means the two are completely different.
Because there is a large covariance between the $\log \Agw$, $\gammagw$, and the $\alpha_i$ parameters, instead of comparing the prior and posterior distributions for the $\alpha_i$, we compare distributions for the power at a given frequency $f_i$. 
We construct convolved priors 
using the posteriors for $\log \Agw$ and $\gamma$ and the prior on $\alpha_i$, then compute the Hellinger distance between these distributions and the posteriors on the power at $f_i$, according to Eq.~\eqref{t-process}.

\subsection{Pure power law background}

\begin{figure}[t!]
    \vspace{-0.2in}
    \centering
    \includegraphics[width=0.75\columnwidth]{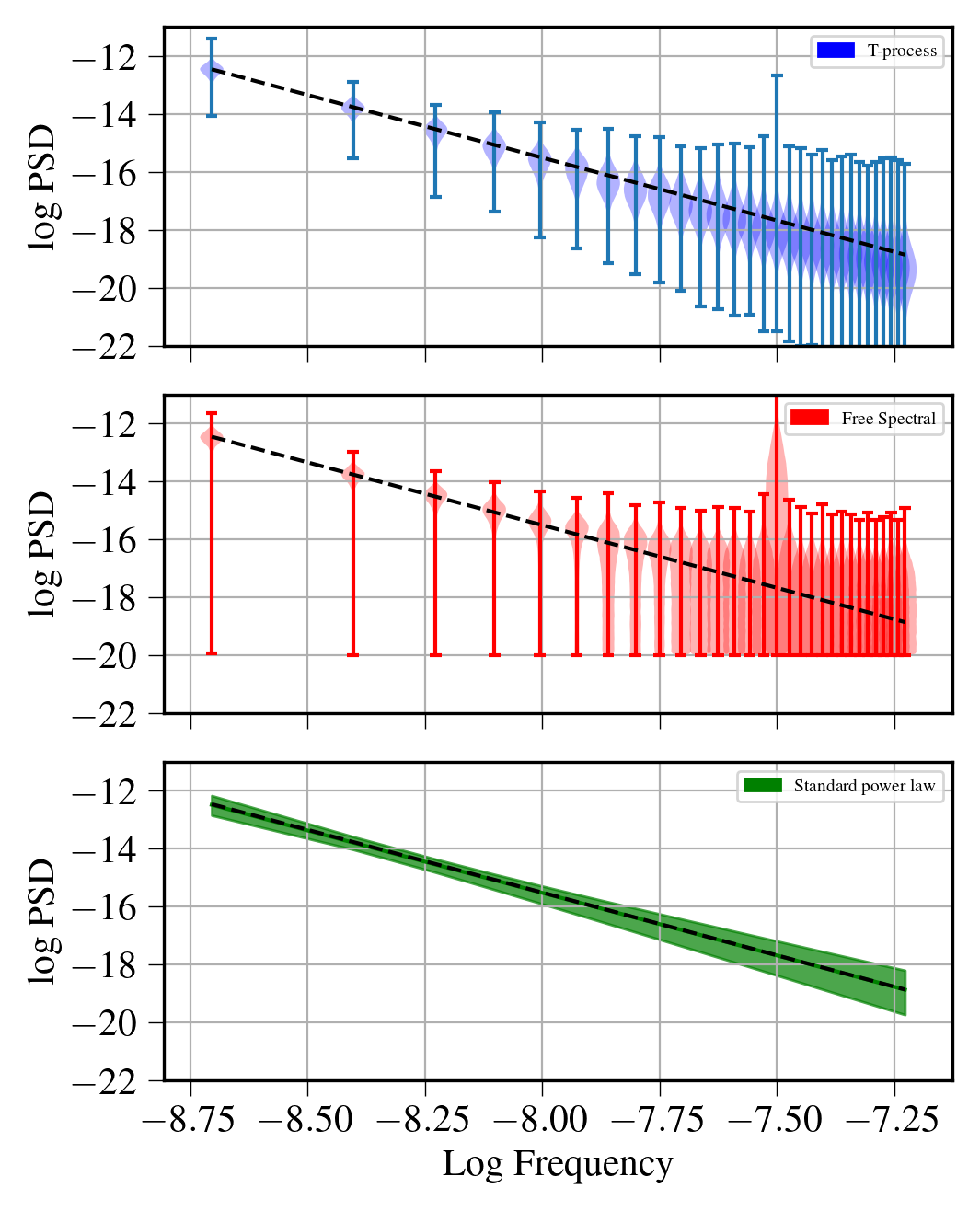}
    \caption{Reconstructed GWB PSD for power-law simulations using a $t$-process model (blue), free spectral model (red), and power-law model (green). The violins and shaded region show the 90\% confidence interval. The injected PSD is shown as the black dashed line. The standard power-law recovers the injected PSD the best, although all three models recover the injected PSD accurately at low frequencies.}
    \label{fig:plaw sim psd}
\end{figure}

\begin{figure}
    \centering
    \includegraphics[width=\columnwidth]{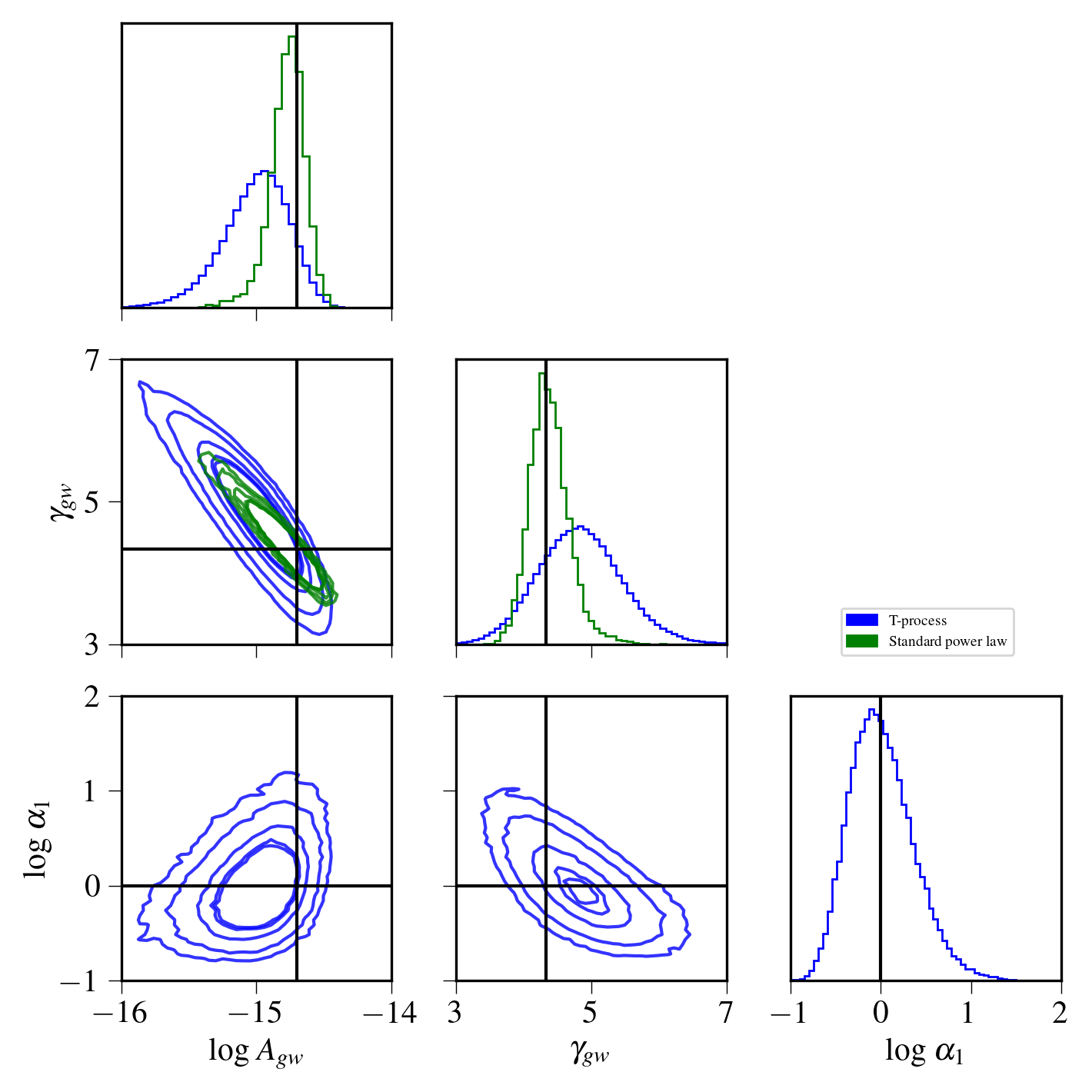}
    \caption{Combined posterior plot for all 50 simulations of $\log \Agw$, $\gammagw$ and $\log \alpha_1$  for the power-law recovery (blue) and $t$-process recovery (green). We see that the standard power law recovers the injected amplitude and spectral index value very accurately. The $t$-process is also able to recover the injected values, but underestimates the amplitude and overestimates the spectral index. We also see that $\log \alpha_1$ is centered at 0, implying there is no deviation from a power-law PSD at that frequency. The other $\alpha_i$ values not shown have similar distributions, with increasing variance as we go to higher frequencies}
    \label{fig:logA+gamma pplaw}
\end{figure}

\begin{figure}[h!]
    \centering
    \includegraphics[width=0.5\columnwidth]{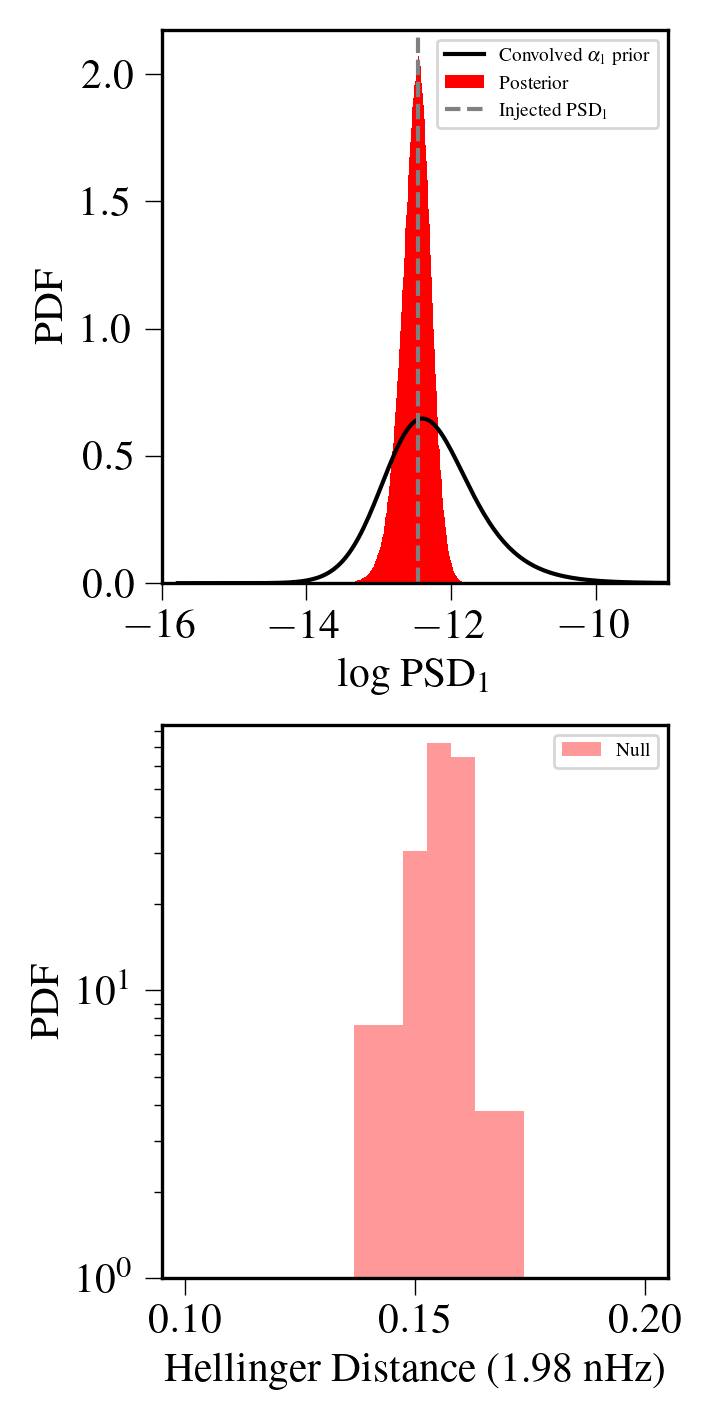}
    \caption{The log-PSD values (top) and the distribution of Hellinger distances for the first frequency (1.98~nHz) of the pure power law injection (bottom) of all 50 simulations. The black curve indicates the convolution of the $\log \Agw$, $\gammagw$ posteriors and the inverse gamma prior. We see that the PSD is accurately recovered at the maximum PDF of the convoluted prior, implying that there is no adjustment of the $\alpha_1$ parameter. This is the null distribution of Hellinger distances (no deviations from the power law) and at 1.98~nHz the 90\% confidence interval is between 0.146 and 0.163.}
    \label{fig:HD null}
\end{figure}

We generate simulated data sets containing a GWB described by a power-law PSD with an amplitude of $\Agw = 2 \times 10^{-15}$ and a spectral index of $\gammagw=13/3$.
This simulation can be thought of as a litmus test to see how the $t$-process model recovers the PSD when there are no deviations present. For an ideal recovery, the $t$-process would find the injected amplitude and spectral index, and we expect the $\log\alpha_i$ posteriors to have maximum PDF values at 0 $\forall f_i$.

In Fig. \ref{fig:plaw sim psd} we present the reconstructed PSD at each frequency component for all three models. We find good agreement between 
the reconstructed PSDs using all three models at low-frequencies: at higher 
frequencies, the reconstructed PSD using the free spectral model has much 
larger variances compared to the other two because the reconstruction is 
dominated by pulsar white noise. By design, the free spectral model does 
not assume any relationship between power at different frequencies, whereas 
the power-law and $t$-process models do. In Fig. \ref{fig:logA+gamma 
pplaw}, we present the posteriors of recovered parameters for the power-law 
and $t$-process models. We find that the power-law model best recovers the 
injected values of $\log \Agw$ and $\gamma$, while the $t$-process slightly 
underestimates the amplitude and slightly overestimates the spectral index. 
We also see that the posteriors of $\log \alpha_1$ are recovered with 
maximum posterior values at 0, which is consistent with a pure power-law 
PSD injection. For the other components, we also see a $\log \alpha_i$ 
distribution that is centered close to 0, and as we go to higher 
frequencies, the variance of the distribution increases.

In Fig. \ref{fig:HD null} we present the posterior of the log PSD (top) and the distribution of Hellinger distances (bottom) of the first frequency component (1.98 nHz) for all 50 simulations. Across these simulations, the Hellinger distances range from 0.146 to 0.163 (90\% confidence interval), indicating that the posteriors are very similar to the priors. 

Since this is the pure power-law simulation, the Hellinger distances we calculate can be treated as a ``null distribution,'' i.e., the distribution of the Hellinger distances in the absence of deviations from a power-law. We use these results as a metric for determining the significance of deviations from a power-law when analyzing our other simulations.

Our sensitivity to deviations varies as a function of frequency. The range of Hellinger distances shown in Fig.~\ref{fig:HD null} are typical for low frequencies up to $\sim$ 13.8~nHz. As we move to higher frequencies, the signal is contaminated by pulsar white noise, resulting in smaller Hellinger distances since the recovered posteriors are more similar to the priors. The lowest Hellinger distance occurs at $16/T_\mathrm{span} \sim 1\mathrm{yr}^{-1}$ where we are least sensitive to GWs due to the Earth's orbital period.

\subsection{Excess power at single frequency}

\begin{figure}[h]
    \centering
    \includegraphics[width=0.75\columnwidth]{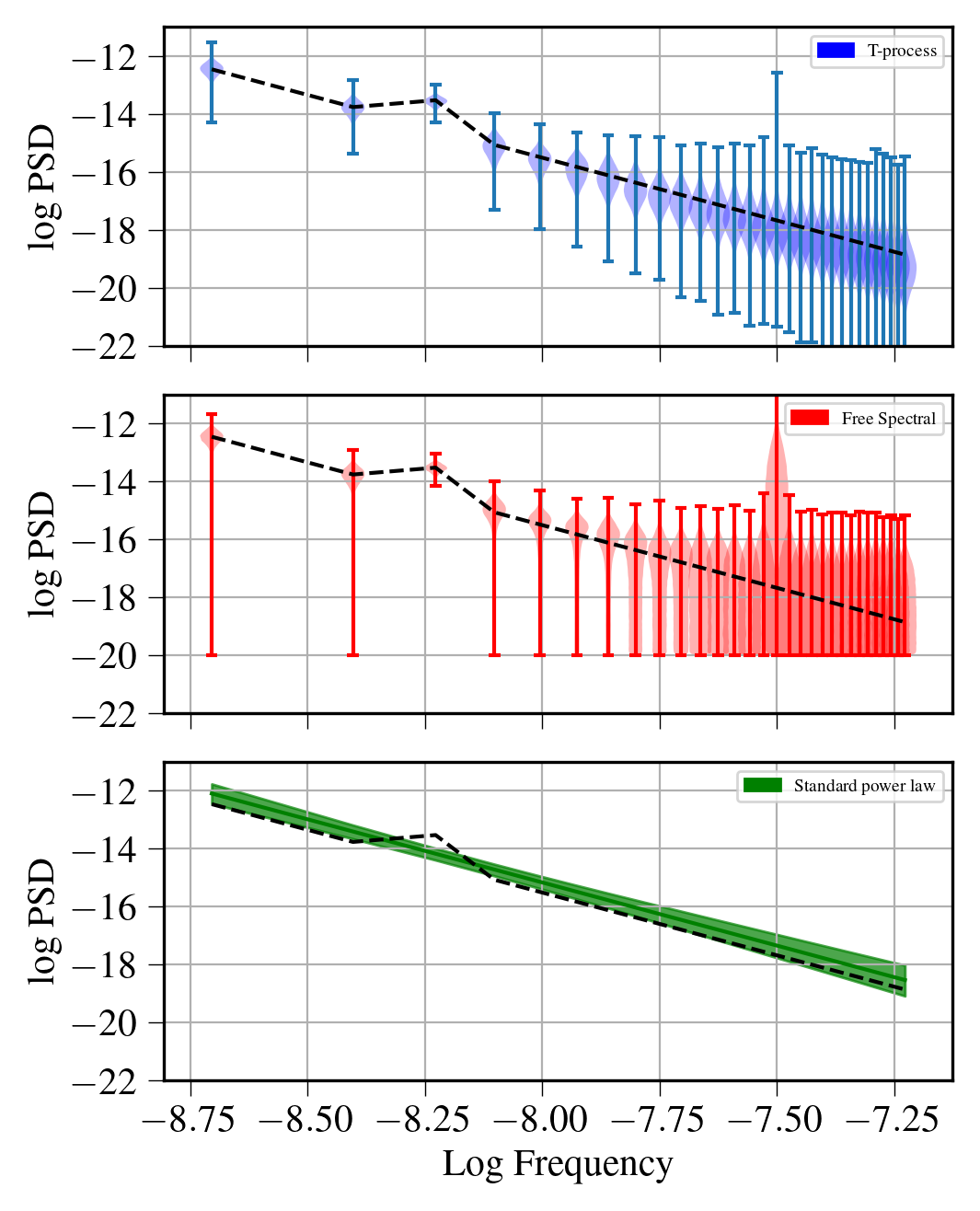}
    \caption{Reconstructed GWB PSD for excess power simulation using a $t$-process model (blue), free spectral model (red), and power-law model (green). The violins and shaded region show the 90\% confidence interval. The injected is shown in the black dashed line. For these simulations, the PSD follows a power-law PSD with excess power at 5.93 nHz. The $t$-process and the free spectral are able to recover the bump as well as the rest of the injection, whereas the standard power law overestimates the amplitude of the entire PSD.}
    \label{fig:3f bump}
\end{figure}

\begin{figure}
    \centering
    \includegraphics[width=\columnwidth]{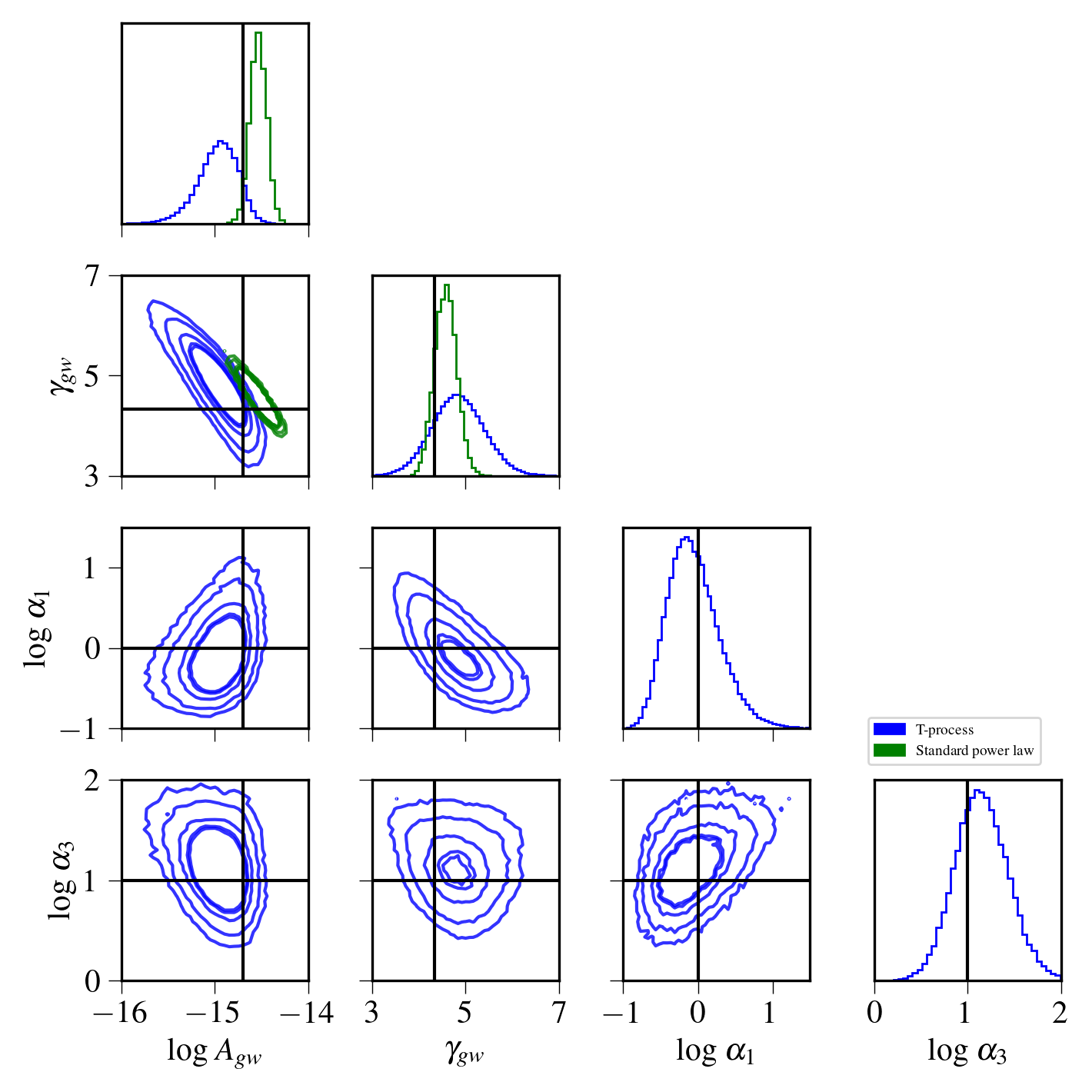}
    \caption{Corner plot of all 50 simulations of the $3^{rd}$ frequency bump simulations. The blue contours and histograms are for the $t$-process, while the green histograms and contours are for the standard power law recovery. We can clearly see that the standard power law model is completely skewed towards higher logA values, whereas the $t$-process captures the injected values, as well as recovers the deviation at $5.93$~nHz. We see that since $\log \alpha_3 = 1.0$, the $t$-process detects the presence of excess power. The other $\log \alpha_i$ distributions not shown here all peak around 0, and as we increase frequency, the variance increases as well.}
    \label{fig:logA gamma 3f bump}
\end{figure}

\begin{figure}[h!]
    \centering
    \includegraphics[width=0.5\columnwidth]{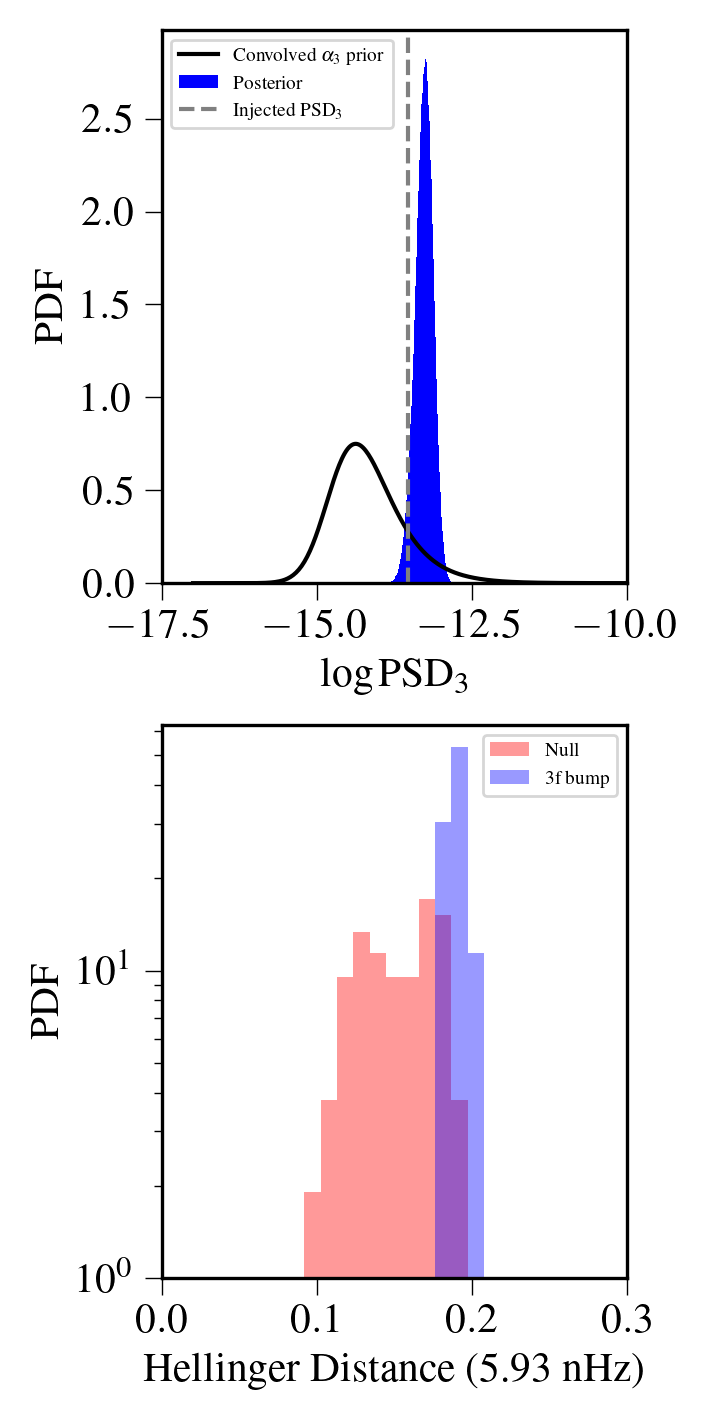}
    \caption{The recovered log-PSD values (top) and the distribution of Hellinger distances (bottom) comparing the 3f bump simulations in blue to the pure power law simulations (null hypothesis) in red of all 50 simulations. The black curve indicates the convolution of the $\log \Agw$, $\gammagw$ posteriors and the inverse gamma prior. The posterior is overestimated from the injected value but still presents a clear distinction from the prior, implying that the $\alpha_3$ parameter is adjusted for the excess noise at 5.93~nHz. The null distribution has a 90\% confidence interval between 0.114 and 0.184, while the excess noise simulations have a 90\% confidence interval between 0.182 and 0.201 at 5.93~nHz.}
    \label{fig:3f bump hellinger}
\end{figure}

In order to test how well we can characterize deviations from a power-law PSD, we generate simulations where the GWB follows a power-law PSD with a known value of excess power at a single frequency. 
Such a feature could be produced by a single nearby circular individual SMBHB, although the detection of such a source would require performing a search using the appropriate deterministic signal model. 

A search of the NANOGrav 15-year data set found no evidence of GWs from individual SMBHBs in \citep{2023ApJ...956L...3A}, but with longer baselines and more pulsars, we expect the sensitivity of PTAs to improve to the point where a handful of sources will be detectable \citep{2015MNRAS.451.2417R,2017NatAs...1..886M,2022ApJ...941..119B}. Excess correlated noise could also be produced by common sources, such as error in the Solar System ephemeris \citep{2012MNRAS.427.2780H, 2019ApJ...876...55R, 2020ApJ...893..112V}, as well as contributions from Solar wind \citep{2021A&A...647A..84T}.

We generate the simulated data sets with a common process with an amplitude of $2 \times 10^{-15}$ and a spectral index of $13/3$, then multiply the power at $5.93$~nHz (the third frequency component in our Fourier basis) by a factor of 10. In terms of the $t$-process model, this corresponds to $\log\alpha_3 = 1$ and $\log\alpha_i = 0$ for $i \neq 3$. This excess noise could be caused by a feature of the background, or an unmodeled red noise process. An important note here is the fact that the source of excess noise will likely be at a frequency different than the bin center. PTAs have a finite frequency resolution in Fourier space ($\Delta f = T_\mathrm{span}^{-1}$). Any excess power within the frequency bin edges will be recovered at the bin centers our PTA is sensitive to. When excess power is in between two Fourier bins, both bins see an excess of power. In this toy-model case, we have chosen to inject the excess power at the bin center so that the only one $\alpha$ parameter sees the deviation. We consider the case where multiple Fourier bins are affected in Section \ref{astropop}.

In Fig. \ref{fig:3f bump} we present the reconstructed PSDs for each of the three models for all 50 simulations. We find that the $t$-process model accurately recovers the injected PSD. The free spectral model accurately recovers the PSD at low frequencies, but at high frequencies the spread in the recovered power is significant due to the presence of pulsar white noise. The power-law model is not able to accurately recover the injected PSD because it cannot recover the deviation at 5.93 nHz. 

In Fig. \ref{fig:logA gamma 3f bump} we present the posterior distributions for the parameters of the $t$-process and power-law models. For the $t$-process model, we find that the amplitude is slightly underestimated while the spectral index is slightly overestimated, as was the case for the power-law simulation (Fig. \ref{fig:logA+gamma pplaw}). We also find that the posterior for $\log\alpha_3$ has a maximum PDF value at 1, indicating that the magnitude of the deviation is accurately recovered, while the posterior at $\log\alpha_1$, where there is no deviation, has a maximum PDF value at 0. In contrast, the power-law model overestimates both the amplitude and spectral index in an attempt to fit the excess power at 5.98 nHz. This shows that the $t$-process can recover deviations without significantly biasing the recovery of $\log \Agw$ and $\gammagw$, unlike the recovery with the power-law model. For all other frequency components, $\log \alpha_i$ peaks close to 0 and the variance of each distribution inceases as we go to higher frequencies.

In Fig. \ref{fig:3f bump hellinger} we present the posterior log PSD (top) and the distribution of Hellinger distances (bottom) at the third frequency component (5.93 nHz) for all 50 simulations (blue histogram). We compare these to the distribution of Hellinger distances from the power-law simulations (red histogram). For the excess power simulations, the Hellinger distances range from 0.182 to 0.201 (90\% confidence interval), whereas for the power-law simulations, the Hellinger distances range from 0.114 to 0.184 (90\% confidence interval). In 82\% of the excess power simulations, the Hellinger distance falls outside the 90\% confidence interval for the power-law simulations. From this, we conclude that we can use the Hellinger distances to determine the significance of deviations from a power-law PSD. We also show the posterior on the recovered power at 5.93 nHz (blue histogram) to the convolved prior (black histogram). We see the posterior differs significantly from the prior, and has a maximum posterior value of -12.674, which is consistent with the injected PSD.

\subsection{Astrophysically realistic GWB from SMBHB populations} \label{astropop}

\begin{figure}[h!]
    \centering
    \includegraphics[width=0.75\columnwidth]{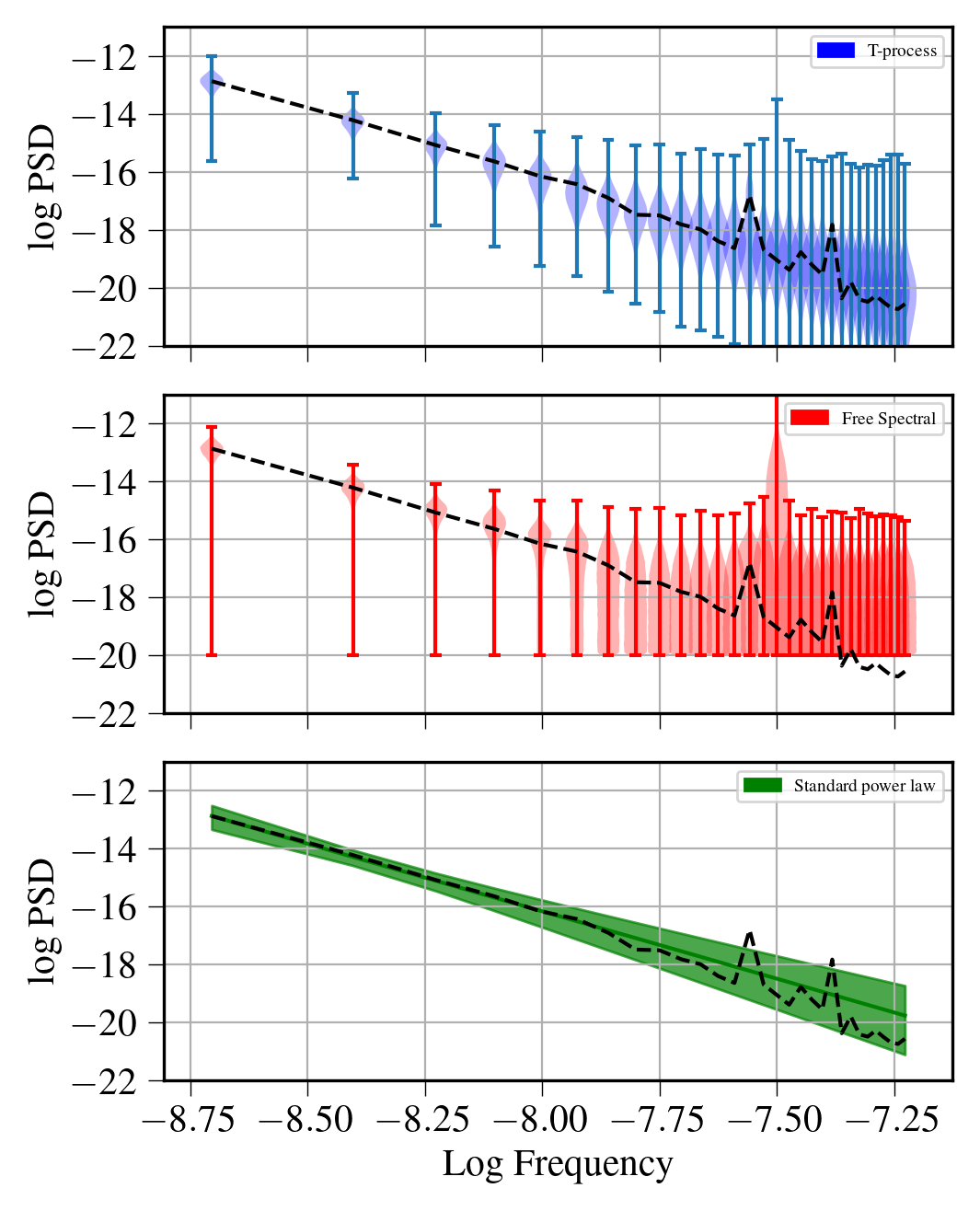}
    \caption{Reconstructed GWB PSD for the \holodeck\ simulations using a $t$-process model (blue), free spectral model (red), and power-law model (green). The violins and shaded region show the 90\% confidence interval. The injected PSD is shown as the black dashed line. The $t$-process model is best able to recover the injected PSD, while the free spectral model and power-law model recover the injected PSD well at low frequencies but not at high frequencies.}
    \label{fig:holodeck sim}
\end{figure}

\begin{figure}
    \centering
    \includegraphics[width=\columnwidth]{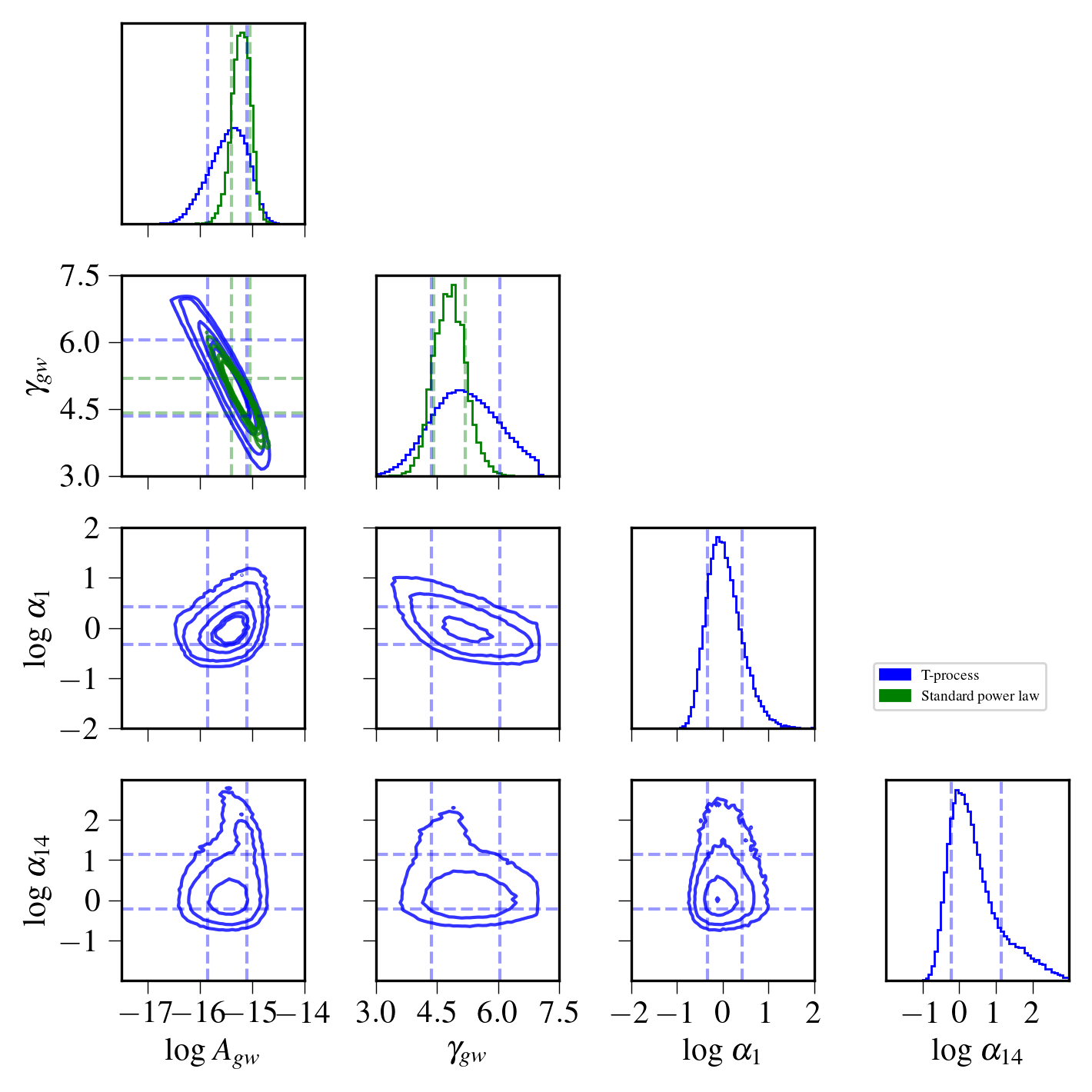}
    \caption{Corner plots of all 50 holodeck simulations. The blue contours and histograms are for the $t$-process, while the green histograms and contours are for the standard power law recovery. The green and blue dashed lines indicate the 1$\sigma$ confidence interval. We see that the both models recovery the same logA and gamma values, but the $t$-process can also detect the presence of a deviation at $27.7$~nHz, which was seen in Fig. \ref{fig:holodeck sim}, whereas the $t$-process cannot detect such a deviation. We also see that $\alpha_{14}$ component searches over more of the inverse gamma prior. The $\log \alpha_i$ distributions not shown here peak between 0.13 and 0.3, with increasing variance as we go to higher frequencies.}
    \label{fig:holodeck logA gamma}
\end{figure}

\begin{figure}[h!]
    \centering
    \includegraphics[width=0.5\columnwidth]{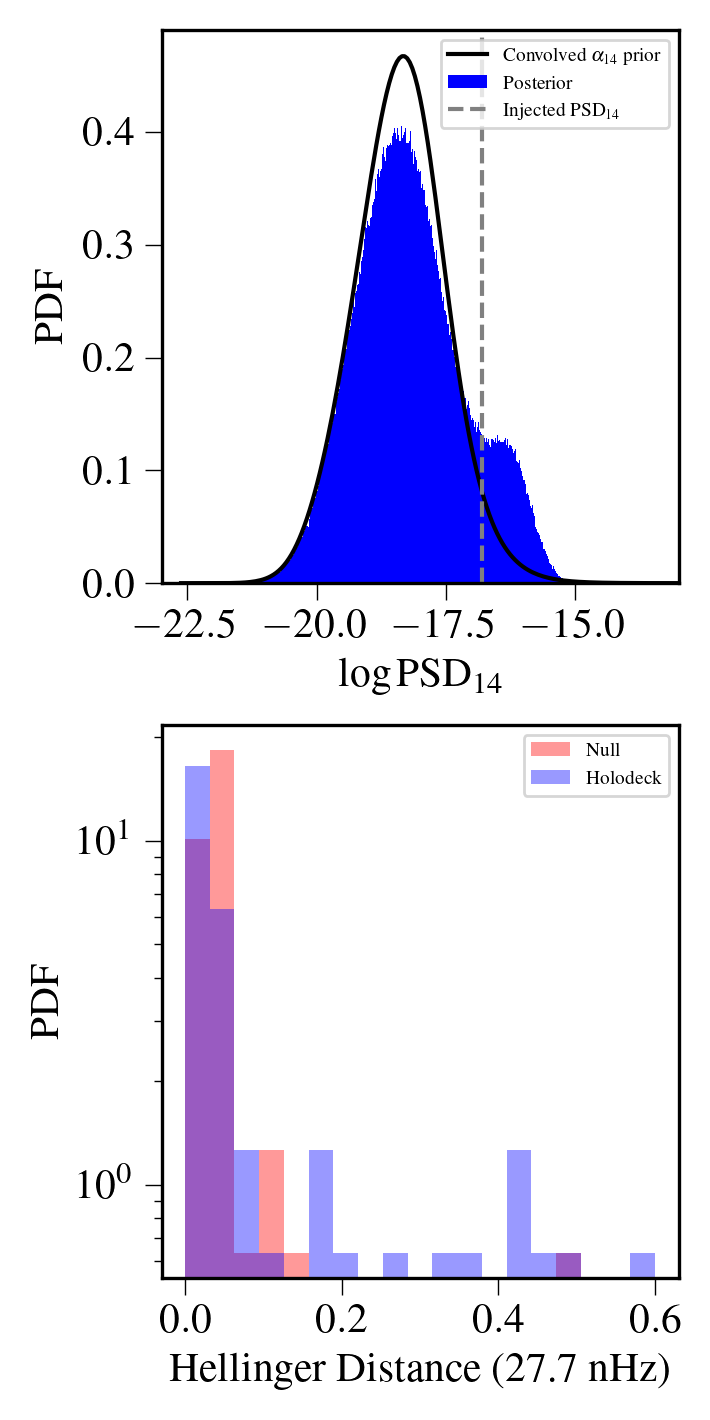}
    \caption{The recovered log PSD (top) and the distribution of Hellinger distances (bottom) comparing all 50 of the holodeck simulations in blue to the pure power law simulations (null hypothesis) in red for the $14^{th}$ frequency component ($27.7$~nHz), where a major deviation from the power law occur. The astrophysically realistic simulations have a 90\% confidence interval between 0.022 and 0.457, whereasthe null distribution has a 90\% confidence interval between 0.027 and 0.104. The bottom panel show the log-PSD posteriors (blue) and the black curve indicates the convolution of the $\log \Agw$, $\gammagw$ posteriors and the inverse gamma prior. At the this frequency component, the posterior tends to search over the entire prior space, but we see a deviation for some simulations, which are able to detect the presence of excess noise at $27.7$~nHz. Simulations with HD $<$ 0.1 are the closest to the prior, whereas simulations with HD $>$ 0.1 display the bimodality of the posterior in the bottom panel.}
    \label{fig:holodeck hellinger}
\end{figure}

In the previous subsections, we have shown how the $t$-process PSD can be used to recover the GWB PSD for two types of simulations: one where the GWB PSD follows a power-law PSD, and one where the PSD follows a power-law PSD with excess power at a single frequency. While these tests show the effectiveness of the $t$-process, it is also important to consider how the $t$-process would analyze realistic PTA data. In this section, we present results from analyzing simulations where we generate the GWB using simulated populations of SMBHBs. 
We use \holodeck\ to generate the GWB PSD, using hyper parameters derived from astrophysical priors as were shown in \citet{2023ApJ...952L..37A}. The values used to generate these simulations are listed in Appendix \ref{injected}. Unlike the previous set of simulations, the GWB measured here is generated from many individual sources that are at frequencies between the fourier bin edges, this leads to a more realistic case where binaries may lie between two different frequency bins contributing to both of them.

In Fig. \ref{fig:holodeck sim} we present the reconstructed PSDs using the $t$-process, free spectral, and power-law models. We see that the $t$-process accurately reconstructs the PSD across all frequencies, including some of the higher frequencies that show deviations from a power-law PSD. 
The free spectral model is able to recover the PSD at low frequencies but not at high frequencies where pulsar white noise dominates. 
The standard power-law is able to recover the PSD at lower frequencies, but fails to capture some of the deviations present at higher frequencies.

In Fig. \ref{fig:holodeck logA gamma} we present the posterior distributions of the parameters for both the $t$-process and power-law models for all 50 simulations. Both the $t$-process and power-law models recover consistent posteriors for $\log \Agw$ and $\gammagw$, although the posteriors from the $t$-process model are significantly broader. 
For the $t$-process model, we also show the posteriors of $\log \alpha_1$ and $\log \alpha_{14}$. 
The $\log \alpha_1$ distribution peaks at 0, implying no deviations, while the $\log \alpha_{14}$ posterior shows a tail in the posterior that extends to high values, indicating that there is some evidence of excess power at a frequency of $14/T_\mathrm{span}=27.7$ nHz. Across all other $\log \alpha_i$, we observe a peak close between 0.13 and 0.3, and a larger spread as we move to higher frequencies.

In Fig. \ref{fig:holodeck hellinger} we present the posterior for log PSD (top) and the distribution of Hellinger distances (bottom) at 27.7 nHz. While the posterior (blue histogram) and prior (black histogram) are very similar, we note that there is a second peak in the posterior that extends beyond the prior. This comes from the fact that for some simulations the excess power at this frequency can be recovered, leading to a prior that is peaked at higher values, but for others it cannot.
We also show the distribution of Hellinger distances compared to the null hypothesis at 27.7~nHz. For the \holodeck\ simulations, the Hellinger distances range from 0.022 to 0.457 (90\% confidence interval), while for the power-law PSD simulations, the Hellinger distances range from 0.027 to 0.104 (90\% confidence interval). For 24\% of the \holodeck\ simulations, the Hellinger distance is outside of the 90\% confidence interval for the power-law simulations, indicating that the deviation at 27.7 nHz from a power-law is significant. 
The \holodeck\ simulations show additional features in the PSD at higher frequencies; however, the distributions of Hellinger distances at these frequencies are very similar to the distribution of Hellinger distances for the power-law simulations, indicating that these deviations are not significant.

\section{Conclusions} \label{concl}

PTA experiments around the world have published the first evidence of a nanohertz GWB which is broadly consistent with that predicted to be produced by a cosmological population of SMBHBs. 
By measuring the GWB PSD, 
we can learn about the astrophysical or cosmological sources producing it. 
In this paper, we used simulated PTA data to study how a $t$-process PSD can recover the GWB. We considered three types of simulations: (1) a pure power-law PSD, (2) a power-law PSD with excess power at a single frequency component (which could be the result of the background, or an unmodeled red noise process), and (3) an astrophysically motivated background from a simulated population of SMBHBs. We have shown that the $t$-process is able to accurately recover the PSD for all three simulation types. In addition, the $t$-process model provides a way to determine the significance of any deviations from a simple power-law PSD.

We show that in all three different simulation types, the $t$-process is accurately able to reconstruct the PSD of the GWB. In simulation set (1), it was able to recover the PSD, but has some spread in its recovery due to the 30 extra parameters ($\alpha_i$) which complicate the model. In simulation set (2), the $t$-process was able to recover not only the amplitude and spectral index but the excess of power at the third frequency component as well. In simulation set (3), the $t$-process was able to recover the PSD accurately at lower frequencies, and was even able to detect deviations present at the higher frequency spectrum in a few realizations. In contrast, the standard power-law model was only able to accurately recover the PSD for all frequencies in simulation set (1). In simulation set (2), it overestimated the PSD across all frequencies, and was not able to measure some of the subtleties at the higher frequencies in simulation set (3). Across all three simulation types, the free spectral model was able to accurately recover the PSD at lower frequencies, but due to its model agnostic nature, was contaminated by the white noise floor, which significantly reduces the information gained at higher frequencies (as is seen in Figs. \ref{fig:plaw sim psd}, \ref{fig:3f bump}, and \ref{fig:holodeck sim}). Furthermore, we were able to measure the significance of deviations measured from the power-law by computing the Hellinger distances and comparing to the distribution of Hellinger distances for the power-law simulations.

While the observed nanohertz GWB is consistent with predictions for a GWB generated by SMBHBs, there are many other possible cosmological sources. 
An important difference between a GWB produced by SMBHBs and cosmological sources is that a GWB produced by SMBHBs is being generated by a discrete number of individual sources: this causes some degree of anisotropy in the GWB \citep{2023ApJ...956L...3A} as well as affecting the GWB PSD. 
In this paper, we have shown how we can use the $t$-process model to reconstruct the GWB PSD and detect deviations from a power-law. 
With longer timing baselines and better telescopes, PTA sensitivity will improve, allowing for more robust spectral characterization of the common process spectrum. If or when deviations from a power-law PSD arise, the $t$-process model will play a critical role in assessing their significance and allow the community to ultimately assess whether the GWB source is astrophysical or cosmological.

\section{Acknowledgements}
We thank Aaron Johnson for help with generating the simulated data sets with \enterprise. We thank Luke Kelly and Emiko Gardiner for their help in generating the astrophysical realizations using \holodeck. We also thank Gabriel Freedman for useful discussions. 
The authors are members of the NANOGrav collaboration, and this work was supported by NSF Physics Frontiers Center award number 2020265. 
JS is supported by an NSF Astronomy and Astrophysics Postdoctoral Fellowship under award AST-2202388.

\clearpage

\appendix
\section{Injected Parameters} \label{injected}

\begin{table}
    \caption{The astrophysical parameters injected into the PTA for the holodeck simulations. The parameters are taken directly from the NANOGrav 15yr constraints paper, or are randomly sampled from astropysical priors (if given) given in the same paper. The MMBulge relation is from Kormendy \& Ho \citep{2013ARA&A..51..511K}}
    \begin{center}
    \begin{tabular}{|l|c|c|}
    \hline \hline
        Parameters & Injected value & Prior \\
    \hline
        Hard $\tau$ & 3.133 Gyr & $\mathcal{U}(0.1,11)$ \\ 
        Hard $r_0$ & 1000 pc &  -\\ 
        Hard $r_\mathrm{char}$ & 100 pc & -\\
        Hard $\gamma_\mathrm{inner}$ & $-$0.467 & $\mathcal{U}(-1.5,0.5)$ \\
        Hard $\gamma_\mathrm{outter}$ & 2.5 & -\\
    \hline
        GSMF $\log_{10} \phi_0$ & $-$2.140 & $\mathcal{N}(-2.56,0.4)$ \\
        GSMF $\phi_z$ & $-$0.6 & -\\
        GSMF $\log_{10} M_{\mathrm{char} 0}$ & 11.370 & $\mathcal{N}(10.9,0.4)$ \\
        GSMF $M_{\mathrm{char}, z}$ & 0.11 & -\\
        GSMF alpha0 & $-$1.21 & -\\
        GSMF alphaz & $-$0.03 & -\\
    \hline
        GPF frac norm all q & 0.033 & -\\
        GPF $m_{\alpha}$ & 0 & -\\
        GPF $q_{\gamma}$ & 0 & -\\
        GPF $z_{\beta}$ & 1 & -\\
        GPF max frac & 1 & -\\
    \hline
        GMT norm & 1.514 & $\mathcal{U}(0.2,5)$ \\
        GMT $m_{\alpha}$ & 0 & -\\
        GMT $q_{\gamma}$ & $-$1 & -\\
        GMT $z_{\beta}$ & $-$.5 & -\\
    \hline
        MMB $\log_{10} m_{\mathrm{amp}}$ & 8.498 & $\mathcal{N}(8.6,.2)$ \\
        MMB plaw & 1.1 & -\\
        MMB scatter dex & 0.292 & $\mathcal{N}(0.32, 0.15)$ \\
        $f_{\mathrm{bulge}}$ & 0.615 & -\\
    \hline
    \end{tabular}
    \end{center}
    
\label{tab:astro hyper params}
\end{table}

\begin{table}
    \caption{The injected intrinsic red noise parameters (spectral index and log amplitude) for all the simulations, as well as the common process power law that was used in the first two types of simulations.}
    \begin{center}
    \begin{tabular}{|l|c|c||l|c|c|}
    \hline
        $\textbf{Pulsar}$ & $\boldsymbol{\gamma}$ & $\boldsymbol{\log_{10}A}$ & $\textbf{Pulsar}$ & $\boldsymbol{\gamma}$ & $\boldsymbol{\log_{10}A}$ \\
    \hline  \hline
        B1855+09 & 3.718 & $-$14.752 & J1730$-$2304 & 3.007 & $-$15.696\\
    \hline
        B1937+21 & 4.002 & $-$13.563 & J1738+0333 & 3.589 & $-$16.711\\
    \hline
        B1953+29 & 2.283 & $-$12.918 & J1741+1351 & 3.021 & $-$16.609\\
    \hline
        J0023+0923 & 3.095 & $-$16.546 & J1744$-$1134 & 3.090 & $-$16.546\\
    \hline
        J0030+0451 & 3.834 & $-$16.228 & J1745+1017 & 2.370 & $-$11.838\\
    \hline
        J0340+4130 & 3.300 & $-$16.929 & J1747$-$4036 & 2.712 & $-$12.612\\
    \hline
        J0406+3039 & 3.508 & $-$16.359 & J1751$-$2857 & 3.394 & $-$16.617\\
    \hline
        J0437$-$4715 & 3.225 & $-$16.721 & J1802$-$2124 & 1.826 & $-$12.284\\
    \hline
        J0509+0856 & 3.370 & $-$15.807 & J1811$-$2405 & 3.341 & $-$16.867\\
    \hline
        J0557+1551 & 3.325 & $-$16.268 & J1832$-$0836 & 3.286 & $-$16.902\\
    \hline
        J0605+3757 & 3.411 & $-$16.251 & J1843$-$1113 & 3.251 & $-$16.553\\
    \hline
        J0610$-$2100 & 3.386 & $-$12.815 & J1853+1303 & 1.890 & $-$14.393\\
    \hline
        J0613$-$0200 & 2.782 & $-$15.311 & J1903+0327 & 1.603 & $-$12.191\\
    \hline
        J0636+5128 & 3.258 & $-$16.860 & J1909$-$3744 & 3.282 & $-$17.226\\
    \hline
        J0645+5158 & 1.872 & $-$14.376 & J1910+1256 & 3.112 & $-$16.736\\
    \hline
        J0709+0458 & 3.616 & $-$15.617 & J1911+1347 & 3.107 & $-$16.757\\
    \hline
        J0740+6620 & 3.232 & $-$17.087 & J1918$-$0642 & 3.137 & $-$16.886\\
    \hline
        J0931$-$1902 & 3.243 & $-$16.879 & J1923+2515 & 3.247 & $-$17.103\\
    \hline
        J1012+5307 & 0.627 & $-$12.641 & J1944+0907 & 2.550 & $-$14.442\\
    \hline
        J1012$-$4235 & 3.513 & $-$15.997 & J1946+3417 & 1.046 & $-$12.462\\
    \hline
        J1022+1001 & 2.694 & $-$15.176 & J2010$-$1323 & 3.134 & $-$16.934\\
    \hline
        J1024$-$0719 & 3.197 & $-$16.755 & J2017+0603 & 3.269 & $-$17.160\\
    \hline
        J1125+7819 & 3.193 & $-$16.863 & J2033+1734 & 3.371 & $-$16.801\\
    \hline
        J1312+0051 & 3.402 & $-$16.603 & J2043+1711 & 3.238 & $-$17.166\\
    \hline
        J1453+1902 & 3.360 & $-$16.773 & J2124$-$3358 & 3.444 & $-$16.565\\
    \hline
        J1455$-$3330 & 3.231 & $-$16.936 & J2145$-$0750 & 0.781 & $-$12.930\\
    \hline
        J1600$-$3053 & 2.806 & $-$16.507 & J2214+3000 & 3.094 & $-$16.913\\
    \hline
        J1614$-$2230 & 3.292 & $-$17.095 & J2229+2643 & 3.344 & $-$17.017\\
    \hline
        J1630+3734 & 3.386 & $-$16.285 & J2234+0611 & 3.325 & $-$16.138\\
    \hline
        J1640+2224 & 3.265 & $-$17.170 & J2234+0944 & 3.269 & $-$16.826\\
    \hline
        J1643$-$1224 & 1.103 & $-$12.302 & J2302+4442 & 3.349 & $-$16.910\\
    \hline
        J1705$-$1903 & 0.946 & $-$11.929 & J2317+1439 & 3.106 & $-$17.075\\
    \hline
        J1713+0747 & 2.469 & $-$15.950 & J2322+2057 & 3.240 & $-$17.014\\
    \hline
        J1719$-$1438 & 3.380 & $-$16.506 & GWB (sims. 1 and 2) & 4.333 & $-$14.699\\
    \hline
    
    \end{tabular}
    \end{center}
    
\label{tab:rn params}
\end{table}

\bibliography{authors}

\end{document}